\documentclass[preprint,3p]{elsarticle}
\usepackage{amsmath,amssymb}
\usepackage{graphicx}
\usepackage{natbib}

\begin{document}
	
	\begin{frontmatter}
		
		\title{Identity and Self as Physical Signatures of Life in Dictyostelium and Multicellular Systems}
		
		\author{Yehuda Roth}
		\ead{yudroth@gmail.com}
		
		\address{Oranim College, K. Tivon, Israel}
		
		\begin{abstract}
			In previous work we sought to address the fundamental question ``What is life?''. Building on that conceptual foundation, we here integrate traditional criteria for living systems with our recent proposals on physical identity and self, and apply them to concrete biological cases. We suggest that life can be characterized as the preservation of a well-defined identity ``at all costs'' together with the presence of a physically grounded self, and we show how this perspective illuminates the organization and social behavior of Dictyostelium and other multicellular systems.
		\end{abstract}
		
		
	\end{frontmatter}
	
	%%%%%%%%%%%%%%%%%%%%%%%%%%%%%%%%%%%%%%%%
	\section{Introduction: Identity and self as signatures of life}
	%%%%%%%%%%%%%%%%%%%%%%%%%%%%%%%%%%%%%%%%
	
	Living systems are commonly characterized by functional properties such as metabolism, replication, homeostasis, responsiveness, and evolution.\cite{Maturana1980,Luisi2016}
	These lists, however, typically treat ``the organism'' as a given bearer of functions, without deriving its individuality from underlying physical degrees of freedom.\cite{Maturana1980,Deacon2012}
	Here we take a complementary perspective and ask how identity and self can be formulated in physical terms, and how such a formulation can serve as a structural signature of life.\cite{Davies2021,Davies2021physicsbio}
	Rather than starting from a fixed notion of ``the organism'', we examine how different physical systems — from single cells and bacterial colonies to multicellular bodies and amoeboid aggregates — come to count as one or many, and how they protect that status against perturbations.\cite{Davies2021,Davies2021physicsbio}
	
	The guiding idea is to view life as a dynamical tendency to preserve a well-defined identity for as long as possible, not merely as persistence of material constituents.\cite{Roth2026ClassicalAging}
	Identity, in this sense, refers to the organization of degrees of freedom (for example genetic information and its expression) that makes a system recognizably the same over time and under changing conditions.
	Self, in turn, refers to the level of organization at which that identity is both carried and actively defended: the same system that is identified as ``one'' also generates and sustains the rules that protect its own continued existence.
	By focusing on identity and self in this physical way, we aim to distinguish living systems from both static material structures and engineered devices that merely implement externally written programs for stability.
	
	%%%%%%%%%%%%%%%%%%%%%%%%%%%%%%%%%%%%%%%%
	\section{Facts Supporting the Definition of Identity and Its Preservation at All Costs}
	%%%%%%%%%%%%%%%%%%%%%%%%%%%%%%%%%%%%%%%%
	A classical picture of biological individuality is provided by the ``selfish gene'' view, in which genetic information behaves as if it were acting to ensure its own persistence.\cite{Dawkins1976} In this perspective, an entire organism—or even a lineage—can be seen as a system organized around the conservation of a particular genetic configuration against external perturbations, with cells, tissues, and physiological processes functioning as coordinated mechanisms that collectively defend this genetic identity from noise, damage, and competition.\cite{Dawkins1976} From a purely chemical standpoint, a DNA molecule might be viewed as just one more polymer in a crowded environment, analogous to a segment of a crystal lattice that passively occupies a particular configuration, but in a living system DNA is embedded in a far richer dynamical context, in which replication machinery, repair pathways, and regulatory networks are tuned to monitor and correct deviations from a reference genetic configuration.\cite{GenomeDNABook,DNARepairReview} This means that the state of the DNA molecule is not merely a passive outcome of local interactions but a target of active maintenance processes that continually work to preserve its informational content, so that DNA in a living cell is not ``just another molecule'' but the physical carrier of identity for the system, surrounded by mechanisms that explicitly aim to keep its configuration within a narrow, identity-defining manifold.\cite{GenomeDNABook,IdentityDNA2024} In the terms developed here, these gene- and DNA-centered processes realize the notion of identity preservation at all costs: the organism continually expends free energy and maintains complex regulation to keep its identity-bearing degrees of freedom within a small neighborhood of preferred configurations, even when this is energetically expensive or locally disadvantageous.\cite{Roth2026ClassicalAging}
	
	%%%%%%%%%%%%%%%%%%%%%%%%%%%%%%%%%%%%%%%%
	\section{Engineered systems and the need for self}
	%%%%%%%%%%%%%%%%%%%%%%%%%%%%%%%%%%%%%%%%
	Technological systems can be designed to exhibit sophisticated forms of self-protection.\cite{Andrade2006SelfProtection,Nesfasil2014SelfProtectSurvey}
	For example, one can construct a computer with a particular software configuration and program it to defend itself against viral attacks or unauthorized modifications.\cite{Andrade2006SelfProtection}
	From the outside, such a computer appears to preserve a well-defined ``identity'' and to resist perturbations that threaten its configuration: its software state, data structures, and access privileges are actively guarded against external interference.\cite{Nesfasil2014SelfProtectSurvey}
	Yet this does not make the computer alive in the biological sense.
	The protective behavior is imposed by external design: the rules that govern detection and response to threats are written into the system by its creators, rather than emerging from an internally generated tendency of the system to conserve its own identity.\cite{Sedikides2021IdentityProtection}
	
	This contrast highlights the need for a physical notion of self.
	A system that merely follows externally specified rules for maintaining a configuration does not, by itself, instantiate a self;\cite{Sedikides2021IdentityProtection}
	it is an implementation of someone else's intention.\cite{Sedikides2021IdentityProtection}
	By contrast, a living system can be characterized as one in which the identity-bearing degrees of freedom and the dynamics that protect them are co-defined: the system is both the bearer of identity and the source of the processes that act to preserve that identity.\cite{Roth2026ClassicalAging}
	In the terms proposed here, a physical definition of self must therefore capture this circularity between ``what is being protected'' and ``what performs the protection'', and must allow for an internally generated tendency to preserve identity at all costs, rather than a merely engineered self-protection policy.\cite{Roth2026ClassicalAging}
	
	%%%%%%%%%%%%%%%%%%%%%%%%%%%%%%%%%%%%%%%%
	\section{Coherent physical systems without self: the center-of-mass example}
	%%%%%%%%%%%%%%%%%%%%%%%%%%%%%%%%%%%%%%%%
	In classical mechanics, the center-of-mass construction provides a simple example of coherence between many constituents: a system of $N$ particles with coordinates $q_i$ and masses $m_i$ can be represented by a single collective coordinate
	\[
	Q_q = \sum_{i=1}^N \frac{m_i}{M}\, q_i,
	\]
	so that the individual positions enter only through their contribution to this global degree of freedom.\cite{Roth2026ClassicalCoherence}
	In this description the particles partially lose their individuality, since, within the chosen frame and coarse-graining, the system is tracked as a single object whose motion is encoded by $Q_q$, and thus the center-of-mass coordinate plays the role of an identity-like variable for the ensemble.\cite{Roth2026ClassicalCoherence}
	
	However, this type of coherence does not by itself define a self.
	The choice of which collection of particles to group together, how to coarse-grain them, and in which reference frame to define the center-of-mass is made by the observer rather than by the system itself: the same physical configuration can be partitioned into subsystems with different centers of mass, and nothing in the internal dynamics of the particles privileges one such collective coordinate ``at all costs''.\cite{Goldstein2002,Roth2026ClassicalCoherence}
	In the terms used here, the center-of-mass coordinate exhibits observer-defined coherence with an associated identity, but because the system does not generate or protect this particular identity-bearing degree of freedom from within, it lacks a physical self.
	
	%%%%%%%%%%%%%%%%%%%%%%%%%%%%%%%%%%%%%%%%
	\section{A bacterial example supporting the definition of self}
	%%%%%%%%%%%%%%%%%%%%%%%%%%%%%%%%%%%%%%%%
	A useful biological example that sharpens our definition of self is provided by genetically modified bacteria carrying plasmids.
	A plasmid is a small, usually circular, extrachromosomal DNA molecule that can replicate independently within a bacterial cell and can be used to introduce new genetic traits into an otherwise unchanged bacterial strain.\cite{PlasmidWiki,PlasmidDavidson}
	In this sense, the experimentalist or cloner plays a role anomalous to that of a programmer in a computer system: identity is altered from the outside by imposing a new informational structure on the system.
	
	Suppose that a bacterial strain is forced to carry a plasmid encoding a trait that is nearly neutral to its viability.
	In that case, the new element can often be tolerated for many generations, especially if its expression does not impose a strong metabolic burden.
	By contrast, if the plasmid encodes a trait that significantly reduces growth, survival, or adaptive flexibility, the bacterial system typically does not preserve that imposed identity indefinitely.
	Instead, it tends to eliminate it through plasmid loss, regulatory suppression, or mutations that weaken or remove the harmful function.\cite{BacterialGenetics,AntibioticResistance}
	
	This distinction is important because the externally imposed genetic change is, in both cases, an alteration of identity.
	Yet the bacterial system does not respond to all imposed identity changes equally.
	Neutral or weakly costly changes may persist, whereas changes that undermine viability are actively filtered out by the internal dynamics of the living system.
	These dynamics include DNA surveillance, immune-like defense systems against foreign genetic elements, repair pathways, regulatory responses, and, at the population level, natural selection acting against cells that bear the more harmful configuration.\cite{BacterialImmuneSystem,WeizmannDefenseSystems,PlasmidWiki}
	
	The conceptual lesson is that self should not be identified with identity itself, nor with any externally written program for preserving identity.
	Rather, self is revealed in the system's intrinsic resistance to identity changes that destroy the conditions of its own continued existence.
	The bacterial cell carries an identity, but more importantly, it also embodies processes that distinguish between imposed changes that can be integrated into life and those that work against it.
	For this reason, self may be understood as the internal authorship of identity-preserving dynamics: the same system that bears identity also generates the processes that defend its viability against externally imposed, life-reducing modifications.\cite{BacterialImmuneSystem,WeizmannDefenseSystems}
	
	%%%%%%%%%%%%%%%%%%%%%%%%%%%%%%%%%%%%%%%%
	\section{Colonies versus organisms: an empirical prediction for self}
	%%%%%%%%%%%%%%%%%%%%%%%%%%%%%%%%%%%%%%%%
	Existing studies on bacterial cooperation and multicellularity have shown that microbes can express costly traits that benefit groups, for example by secreting shared protective factors, building extracellular matrices, or forming complex fruiting bodies.\cite{BacterialCooperationReview,BiofilmStructureReview,BacterialMulticellularityReview}
	Such traits can impose substantial fitness costs on individual cells while contributing to the persistence or spread of the collective structure.\cite{BacterialCooperationReview,CooperativeVirulencePlasmid}
	Horizontal transfer of elements can shape community dynamics.\cite{Coscollia2011,Larbig2002}
	However, these experiments usually treat the group either as a colony or implicitly as an organism, without explicitly asking how the same trait behaves when the relevant level of self changes from individual cells to a multicellular body.
	
	In our framework, this distinction becomes central.
	Consider a perturbation (for instance, a genetically encoded infection or stress response) that reduces the survival or growth of each affected cell, but improves the survival of a larger structure, such as a biofilm, a fruiting body, or an amoeboid aggregate, under certain environmental conditions.\cite{BiofilmStructureReview,BacterialMulticellularityReview}
	In a mere colony, where self is realized at the level of individual cells and each cell primarily protects its own viability, we expect the internal dynamics to act against such a perturbation:
	cells will tend to repair, suppress, mutate away, or lose the harmful trait, and natural selection will favor those cells that successfully restore their individual fitness.\cite{BacterialCooperationReview}
	From the standpoint of the colony, the trait is not stabilized; the system behaves as an ensemble of selves rather than as a single self.
	
	By contrast, in a true multicellular organism or a transient amoeboid body, self can be realized at the level of the larger structure.\cite{BacterialMulticellularityReview,ApoptosisOverview}
	Here the same perturbation could be stabilized as a sacrificial mechanism.
	A subset of cells might consistently retain the harmful trait because their reduced individual viability is compensated by increased robustness of the organism as a whole:
	they secrete protective factors, form supportive scaffolds, or undergo programmed death in ways that are beneficial to the collective.\cite{ApoptosisOverview,CellDeathReview}
	From this perspective, what appears to be ``against the nature'' of a cell is entirely natural for the self of the organism.
	The perturbation is not eliminated, but incorporated into the organism's identity-preserving dynamics.
	
	To our knowledge, no experiment has yet been designed to directly test this difference by comparing the fate of a harmful-but-collectively-beneficial trait in a colony versus in an organismal configuration within the same biological system.\cite{BacterialCooperationReview,BacterialMulticellularityReview}
	We therefore regard the following as an empirical prediction of our definition of self:
	a perturbation that damages individual cells but can improve the survival of a higher-level structure will be actively suppressed and removed in a colony, but can be maintained and functionally integrated in a multicellular body, precisely because self shifts from the level of the cell to the level of the organism.
	
	Amoeboid model systems such as the social slime mould \emph{Dictyostelium discoideum} provide a concrete biological example of this duality.\cite{Kessin2001,Schaap2005}
	In nutrient-rich conditions, \emph{Dictyostelium} cells behave as independent amoebae, each functioning as a separate unicellular organism.
	Under starvation, genetically similar cells aggregate to form a transient multicellular body with differentiated cell types, including a sacrificial stalk that supports the dispersal of viable spores.\cite{BacterialMulticellularityReview,Schaap2021}
	Organisms undergo complex spatiotemporal orchestration.\cite{KatohKurasawa2026}
	This makes \emph{Dictyostelium} an ideal system in which to test our prediction:
	the same genetic identity can participate either in a mere colony of unicellular selves or in a multicellular organism whose self is realized at the level of the aggregate, depending on environmental conditions.
	
	%%%%%%%%%%%%%%%%%%%%%%%%
	\section{Summary}
	%%%%%%%%%%%%%%%%%%%%%%%%
	In this work we have treated identity and self as physical signatures of life, rather than as abstract labels attached to organisms.
	A living system is not merely one that happens to maintain a particular configuration, but one in which identity-bearing degrees of freedom (such as genetic information) are both represented by physical states and actively protected by internal dynamics ``at all costs''.
	The selfish gene picture illustrates how entire organisms and lineages can be organized around the preservation of genetic identity, yet engineered systems show that externally imposed identity protection is not sufficient for life: a computer can defend its software configuration while instantiating only the intentions of its designers.
	We therefore propose that life is characterized by a circular coupling between ``what is being protected'' and ``what performs the protection'': the same system that carries identity also generates and sustains the rules that maintain it.
	
	Within this perspective, self is not an extra psychological or functional feature added on top of biology, but a structural property of living dynamics.
	Self is realized when the system that bears identity also authors and enforces its own identity-preserving processes, resisting externally imposed changes that destroy the conditions of its continued existence.
	This view sharpens the distinction between mere colonies and genuine organisms: in a colony, self is largely realized at the level of individual cells, which tend to eliminate harmful identity changes even if these could benefit the group, whereas in a multicellular organism or an aggregated amoeboid body, self can be realized at the level of the collective, allowing sacrificial traits that harm some components to be integrated as protective mechanisms for the whole.
	Examples from bacteria, amoebae, and multicellular organisms suggest that the same genetic identity can participate in different selves, depending on which level of organization is actively writing and revising the rules of protection.
	
	Accordingly, we suggest that the question ``what is life?'' can be reframed in physical terms as ``what kinds of systems generate and maintain their own identity-preserving dynamics?''.
	A system is alive, in our sense, when it not only occupies a recognizable identity state, but also internally sustains the laws that keep that identity compatible with continued existence.
	This shifts the focus from lists of biological functions to the organization of self: the level at which a system counts as one, and the way in which it protects that unity against perturbations, whether in a single cell, a colony, or a multicellular organism.

\end{document}